\documentclass[12pt]{article}  
\usepackage{amsmath}
\usepackage{url}
\usepackage{graphicx}
\usepackage{lscape}
\usepackage{rotating}
\usepackage{epstopdf}
\usepackage{setspace} 
\usepackage[font=footnotesize]{caption}
\usepackage{overcite}

\newenvironment{sciabstract}{%
\begin{quote} \bf}
{\end{quote}}
\usepackage[pagewise]{lineno}

\newcounter{lastnote}

\title{Matching-centrality decomposition and the forecasting of new links in networks}
\author{Rudolf P. Rohr$^{1,2}$\footnote{To whom correspondence should be addressed. E-mail: rudolf.rohr@ebd.csic.es} , Russel E. Naisbit$^1$, Christian Mazza$^3$, \\ and Louix-F\'elix Bersier$^1$. \\
\\        
\\$^1$ Unit of Ecology and Evolution, Department of Biology, \\ University of Fribourg, \\ Chemin du Musée 10, 1700 Fribourg, Switzerland.
\\$^2$ Integrative Ecology Group ,\\ Estaci\'on Biol\'ogica de Do\~nana, EBD-CSIC, \\ Calle Am\'erico Vespucio s/n, 41092 Sevilla, Spain.
\\$^3$ Department of Mathematics, \\ University of Fribourg, \\ Chemin du Musée 23, 1700 Fribourg, Switzerland.}

\date{}

\begin{document}
\maketitle

\newpage

\begin{sciabstract}
Networks play a prominent role in the study of complex systems of interacting entities in biology, sociology, and economics. Despite this diversity, we demonstrate here that a statistical model decomposing networks into \textit{matching} and \textit{centrality} components provides a comprehensive and unifying quantification of their architecture. First we show, for a diverse set of networks, that this decomposition provides an extremely tight fit to observed networks. Consequently, the model allows very accurate prediction of missing links in partially known networks. Second, when node characteristics are known, we show how the \textit{matching-centrality} decomposition can be related to this external information. Consequently, it offers a simple and versatile tool to explore how node characteristics explain network architecture. Finally, we demonstrate the efficiency and flexibility of the model to forecast the links that a novel node would create if it were to join an existing network.\\
\end{sciabstract}

\section*{Introduction}

The modern world is an increasingly connected place, through transport, social and economic networks, and via our knowledge of interactions at the ecological or molecular level \cite{Cohen1978,Maslov2002,Newman2010}. It is increasingly recognized that such systems should be studied globally, and networks of interacting entities provide a powerful representation of their structure and function. Research on network theory parallels this growth \cite{Newman2010}. A first body of research concentrates on the fact that observed networks are often considered to be only partially known. This would be the case in a food web, for instance, in which some real interactions may have yet to be observed, or a protein interaction network where not all pairwise combinations had been tested in the laboratory. Thus, observed links are typically considered as certain, while an absence of a link between a pair of nodes may reflect an absence of information rather than a real absence of interaction. Models have been devised to predict these \textquotedblleft missing links \textquotedblright and thus correct a network dataset for this incomplete sampling or direct future research towards these candidate interactions \cite{Clauset2008,Guimera2009}.\\

A second domain aims to determine if the structure of these networks exhibits basic generalities, and to uncover the processes that may generate these patterns. This aspect has been tackled with a variety of mostly comparative approaches, such as those treating the classification of networks \cite{Guimera2007,Onnela2012}, motifs \cite{Milo2002}, or stochastic models \cite{Barabasi1999}. Progress in this undertaking could be achieved if there were general methods to relate network structure to characteristics of the nodes. For example, body size has been related to patterns in food webs \cite{Petchey2008}, or country politics and trade to the organisation of military conflict networks \cite{Ward2007}.\\

A third potential application of research is network forecasting, which would make network theory a predictive science. Many current issues facing human society would benefit from the ability to forecast networks for new nodes joining them, such as for the ecological interactions of invasive species \cite{Memmott2009}, the molecular interactions of a newly discovered protein \cite{Garcia-Garcia2012}, or the links of a subversive social group \cite{Krebs2002}. However, there exists no general framework for this forecasting. Here we provide a general model that can be applied to all three domains of network research.\\

The key feature of our methodology is the development of a model for the probability of interaction between nodes based on the decomposition of network architecture into \textit{matching} and \textit{centrality} terms. The \textit{matching} term quantifies assortative structure in who makes links with whom \cite{Rossberg2006,Rossberg2010}, while the \textit{centrality} term captures variation in the number of links that nodes make. Typically, research on network structure has focused on patterns in either assortativeness or centrality. However, the architecture of empirical networks is usually a product of both features simultaneously (see e.g. Fortuna et al. \cite{Fortuna2009}). Here, we take into consideration both patterns. Specifically, the decomposition is implemented at the node level, with each node characterized by latent traits of \textit{matching} and \textit{centrality}. Latent traits are variables whose values are unknown a priori, but can be estimated a posteriori from the network adjacency matrix itself \cite{Hoff2002,Rohr2010}. The model, called the \textit{matching-centrality} model, is implemented in such a way that the closer the \textit{matching} traits of two nodes, the greater the probability that they are linked, and the higher the \textit{centrality} trait of a node, the greater the probability that this node makes links.\\

First, based on a dataset of seven networks from disparate fields, we find that this decomposition provides a very precise fit to observed networks. As a result, the model can be used to very accurately predict missing links, as we demonstrate for a terrorist association network. Second, we show that the latent traits of \textit{matching} and \textit{centrality} are not just abstract traits, but can be linked to external information about the nodes and thus provide a means to study network organisation. For example, in an ecological network of trophic interactions, the latent traits are related to the body size and the phylogeny of the interacting species. Finally, by placing latent traits as intermediates between the network structure and the characteristics of the nodes, the model offers the possibility to forecast the interactions made by novel nodes when joining the network. For example, in a spatial network of mammal communities on mountains, we show that we can accurately predict the mammal fauna of unsampled mountains based on geographical characteristics.\\

\section*{Results}

\subsection*{The \textit{matching-centrality} model}

\begin{figure}[t]
	\centerline{\includegraphics*[width= 0.9 \linewidth]{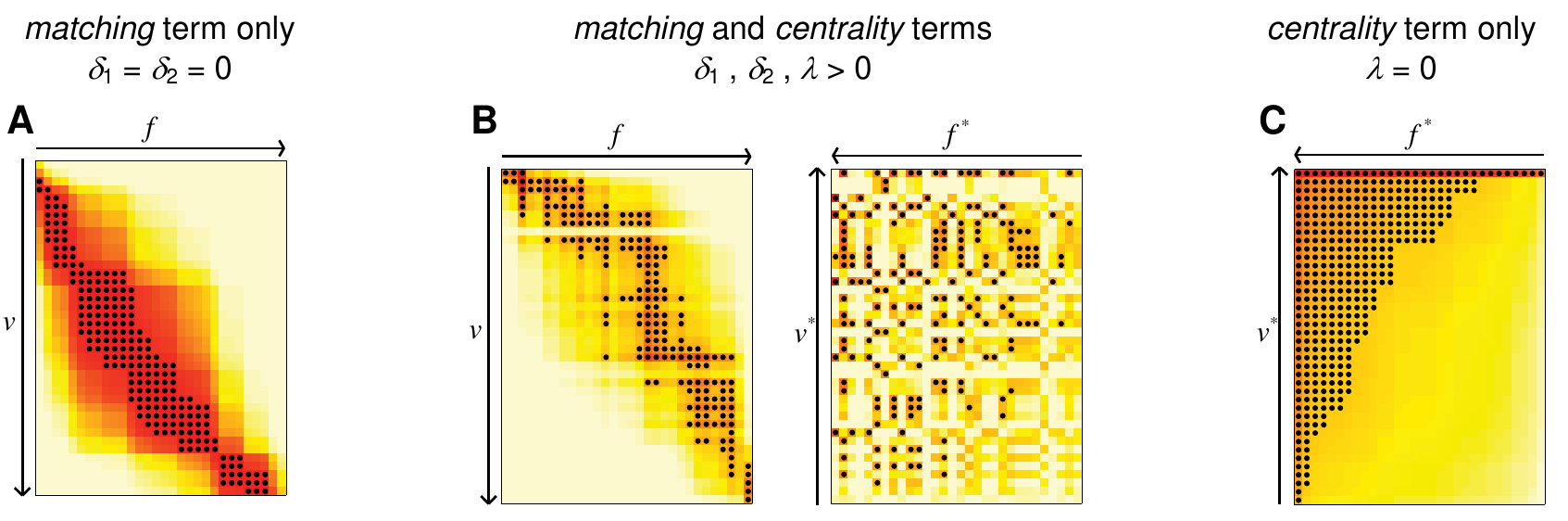}}
	\scriptsize
	\caption{The \textit{matching-centrality} model. The probability of a link between two nodes is decomposed into a \textit{matching} term, quantifying assortative structure in who makes links with whom (A), and a \textit{centrality} term, capturing the fact that nodes can vary considerably in their degree (C). The aim is to simultaneously quantify latent traits of the nodes that are responsible for the \textit{matching} (v, f) and the \textit{centrality} (v*, f*) in the network (B). Panels A, B and C show adjacency matrices of three simulated networks, where a black dot represents a link and colors, from yellow to red, represent increasing linking probability computed with the model. The nodes of the matrices are ordered according to their \textit{matching} or \textit{centrality} traits. If a network exhibits a modular structure \cite{Melian2004,Newman2004,Newman2010}, this will be captured by the formation of clusters in the \textit{matching} traits, while variation in node degree \cite{Price1965,Barabasi1999,Amaral2000,Newman2010} is captured by the \textit{centrality} traits.}
	\label{fig1}
\end{figure}

The model is formulated for undirected bipartite networks, but can be applied to any kind of undirected or directed network, as explained below. Bipartite networks are made of two sets of nodes ($S_1$ and $S_2$) with connections only between them and not within; plant-pollinator networks provide a classical example. Let   be the adjacency matrix of the network, i.e., $a_{ij} = 1$ if there is a link between nodes $i$ and $j$, and zero otherwise. The model characterises each node $i$ in set $S_1$ by a latent trait of \textit{centrality} denoted $v_i^*$, and by $d \geq 1$ latent traits of \textit{matching} \cite{Rossberg2006,Rossberg2010} denoted $v_i^1,\dots ,v_i^d$, and similarly, each node $j$ in set $S_2$ by a centrality trait $f_j^*$ and matching traits $f_j^1, \dots , f_j^d$. The value of $d$ gives the number of matching space dimensions and can be tuned to improve the goodness-of-fit of the model. We take a statistical approach, in which the probability of existence of a link between a pair of nodes $i$ and $j$ (hereafter the linking probability $P(a_{ij} = 1)$ is modelled through its logit \cite{Kolaczyk2009}. Our model is given by
\begin{equation}
\label{equ:1}
	\log\left(\frac{P(a_{ij} = 1)}{1-P(a_{ij} = 1)}\right) = \underbrace{- \sum_{k=1}^{d} \lambda_k (v_i^k - f_j^k)^2}_{\text{matching term}} + \underbrace{\delta_1 v_i^* + \delta_2 f_j^*}_{\text{centrality term }} + m
\end{equation}
where $\delta_1, \delta_2, \lambda_1, \dots , \lambda_d$ are positive constants that scale the relative importance of the \textit{matching} and \textit{centrality} terms and $m$ the common intercept (Methods). For a given dimension $d$, the model parameters and latent trait values for each node can be estimated using a simulated annealing algorithm (Methods). Fig. \ref{fig1} depicts patterns in interaction networks that the model is able to capture. Application to other types of network requires simple modifications: for directed unipartite networks (e.g., food-web or military conflict networks) the two sets of nodes are identical ($S_1 = S_2$); for undirected unipartite networks like most social networks, the adjacency matrix is symmetric, so we have to impose $v_i^k = f_i^k$ , $v_i^* = f_i^*$, $\delta_1 = \delta_2$ and also that the probability of a self-link is equal to zero ($P(a_{ii} = 1) = 0$); for more complex networks like directed or undirected multipartite networks, linking probabilities must be set to 0 for pairs of nodes that, by definition, cannot be linked.\\

\subsection*{Performance of the model}

We illustrate the ability of the \textit{matching-centrality} model to capture network architecture on a set of seven examples from disparate fields (described further in the SI): social interactions in Zachary's karate club network \cite{Zachary1977}, associations between the terrorists involved in the September 11 attacks \cite{Krebs2002}, military conflicts between countries \cite{Ward2007}, protein interactions in Saccharomyces cerevisiae \cite{Maslov2002}, the food web of Tuesday Lake \cite{Jonsson2005}, the mutualistic seed-dispersal web of Nava de las Correhuelas \cite{Rezende2007}, and presence-absence data of mammal species on peaks within the southern Rocky Mountains \cite{Patterson1984}. We fit the model for one and two dimensions of \textit{matching} traits and calculate the AUC, i.e., the area under the receiver operating characteristic (ROC) curve \cite{Kolaczyk2009}. The AUC is a standard measure of the performance of a classifier, which can be interpreted as the probability that the linking-probability of a true positive is higher than the linking-probability of a true negative (AUC = 0.5 for a random classifier, and 1 for a perfect classifier). Our results yield, for one matching dimension ($d = 1$), AUC values falling between $0.9702$ and $0.9967$ and between $0.9977$ and $1$ for $d = 2$, three networks being fitted perfectly (Table S1). The ability of the \textit{matching-centrality} model to capture network architecture is thus high.\\

\subsection*{Exploring networks using the\textit{ matching} and \textit{ centrality} traits}

\begin{figure}[t]
	\centerline{\includegraphics*[width= 0.5 \linewidth]{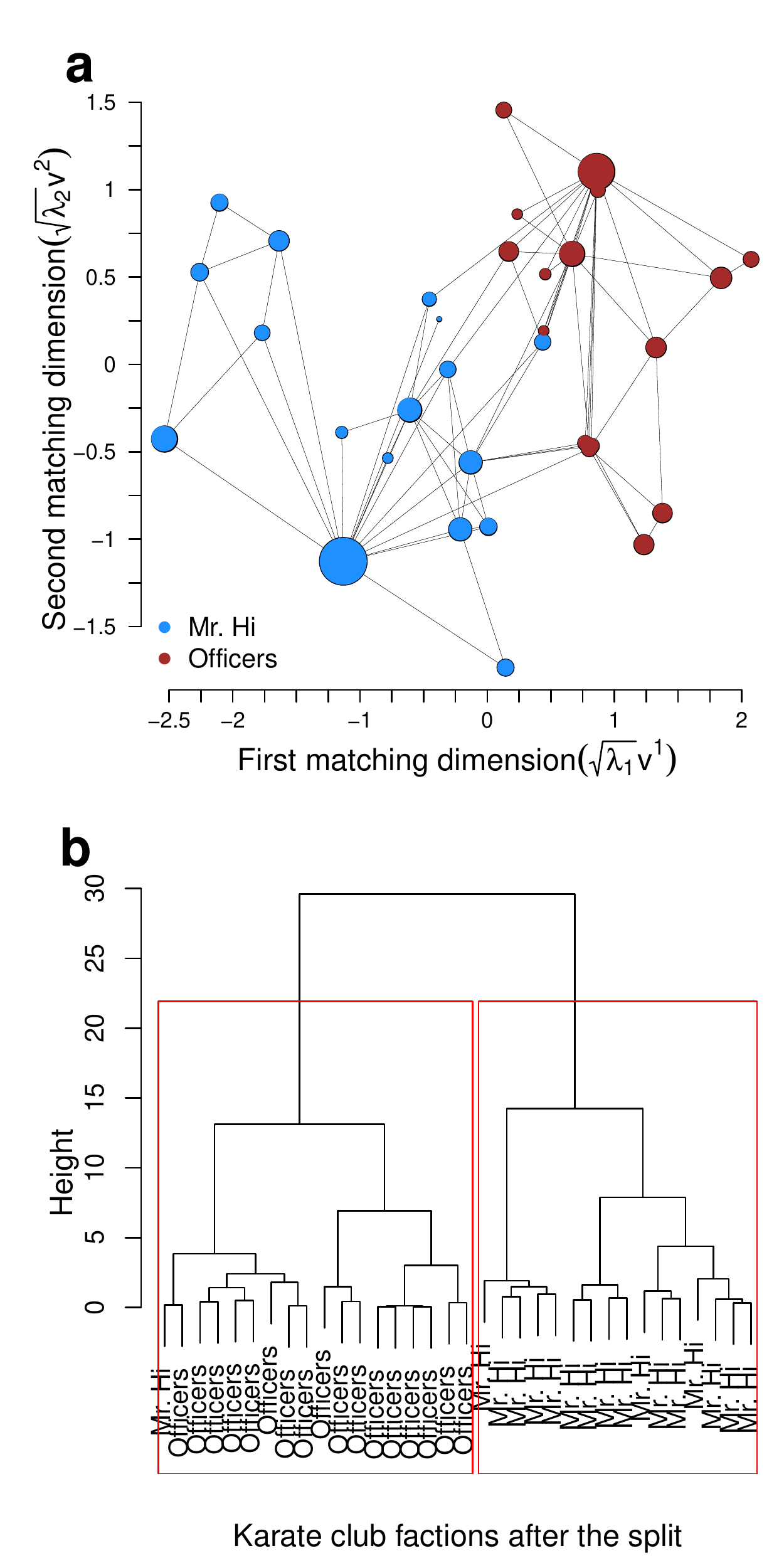}}
	\caption{Panel a shows the two-dimensional \textit{matching} traits space representation of Zachary's karate-club network \cite{Zachary1977}. The size of the nodes (club members linked by friendship) is proportional to their \textit{centrality} trait value. The nodes are displayed in blue or red representing the factions after the split of the club around two members with high \textit{centrality}. Pannel b shows the Ward's clustering of the nodes based on the \textit{matching} and \textit{centrality} traits.}
	\label{fig2}
\end{figure}

Once fitted, the model provides a new representation of the network in the latent trait space (Fig. \ref{fig2}, Fig. S2-S7). For example, the plot of the Zachary karate club network (Fig. \ref{fig2}a), which describe the friendship between the 34 members of a university karate club in the period 1970-1972, exhibits two members that have high \textit{centrality} and divergent \textit{matching} traits. Shortly after the observations, there was an internal dispute and the club split into two factions, denoted \textquotedblleft Mr. Hi \textquotedblright and \textquotedblleft Officers \textquotedblright. The latent traits could have been used to predict the fault line of this fission: the two groups that formed around these two members can be predicted almost perfectly. The Ward's hierarchical clustering, based on the \textit{matching} and \textit{centrality} traits, clearly shows two distinct groups (Fig. \ref{fig2}b), which matches almost perfectly (except for one member) the factions formed when the club split after the dispute (Methods).\\

\subsection*{Prediction of missing links in partially known networks}

\begin{figure}[t]
	\centerline{\includegraphics[width= 0.5 \linewidth]{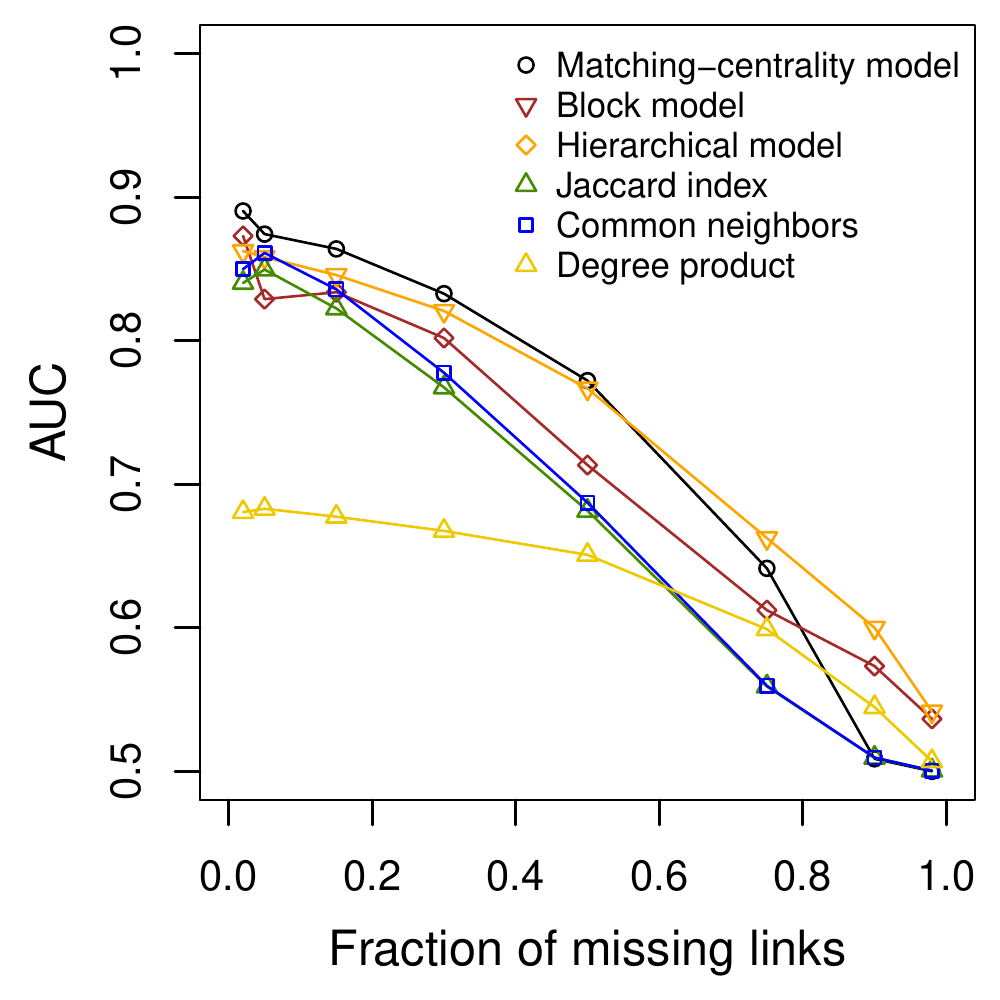}}
	\caption{Prediction of missing links in a partially known network. We present the performance of the \textit{matching-centrality} model (with one dimension) in predicting missing links for the terrorist association network \cite{Krebs2002}, and compare it to several alternative methods \cite{Clauset2008,Guimera2009}. The average AUC statistic (the probability that a missing link is given a higher linking probability than a true negative) is represented as a function of the fraction of simulated missing links created by deleting links in the observed network.}
	\label{fig3}
\end{figure}

In addition to the exploration of network structure, the \textit{matching-centrality} model can be used for the prediction of \textquotedblleft missing \textquotedblright links in partially known networks, where the absence of an interaction may in fact reflect an absence of information \cite{Liben-Nowell2007,Clauset2008,Guimera2009}. Here we demonstrate its performance by simulating missing links in the terrorist association network \cite{Krebs2002}, by setting to 0 a given percentage of links and attempting to recover them (Methods). In the case of this simulation we can judge performance based on the AUC criterion, as the probability that a deleted link is given a higher linking probability than a real absence. For this example, the \textit{matching-centrality} model outperforms other methods \cite{Liben-Nowell2007,Clauset2008,Guimera2009} in recovering these deleted links (Fig. \ref{fig3}). However, when more than $50\%$ of the information is missing, the hierarchical decomposition \cite{Clauset2008} is more precise. This pattern can be explained by the fact that, with a low fraction of observed links, our model tends to overfit the remaining network, and thus underestimates the linking probability for missing links. In practical applications, the absent links with the highest predicted linking probability would be considered to be missing links. These are the candidate interactions that should come under the scrutiny of researchers, thus serving as a guide for cost-effective analysis of complex systems. Unlike previous methods \cite{Clauset2008,Guimera2009}, our approach is not limited to undirected networks.\\

\subsection*{Linking latent traits and node characteristics}

As shown above, the model can be fitted and used for prediction simply based on the network itself. However, it also offers an intuitive tool to gain insight into the processes underlying network structure, as the \textit{matching} and \textit{centrality} traits of nodes can be related to independent information about them using standard analyses such as linear models or Mantel tests. In the food web of Tuesday Lake \cite{Jonsson2005,Naisbit2012}, both the \textit{matching} and \textit{centrality} traits of predators and prey are related to their body size and phylogeny (Methods, Table S2). The latent traits are thus not just an abstract characterisation of the nodes, but provide a versatile method to unravel factors underlying the different aspects of network structure.\\

\subsection*{Forecasting the links of new nodes}

Finally, a significant feature of the \textit{matching-centrality} model is the possibility to forecast the links that new nodes would create when joining an existing network. This might be applied, for example, to forecast the interactions of an invasive species entering a food web or pollination network, the contacts of a non-surveyed individual in a terrorist network, or the biota of an unsampled mountain. The procedure is as follows: from the adjacency matrix we first estimate the latent traits of \textit{matching} and \textit{centrality} for each node in the existing network, and verify that our model provides an accurate fit. Then, using appropriate statistical models, we relate the latent traits of the nodes to external information about them, and ensure that the latter provides a good predictive power for the former. If both conditions are met, we can predict the \textit{matching} and \textit{centrality} traits of the new node(s) using the external information, and finally their linking probability with each of the existing nodes. 

\begin{figure}[t]
	\centerline{\includegraphics[width= 0.5 \linewidth]{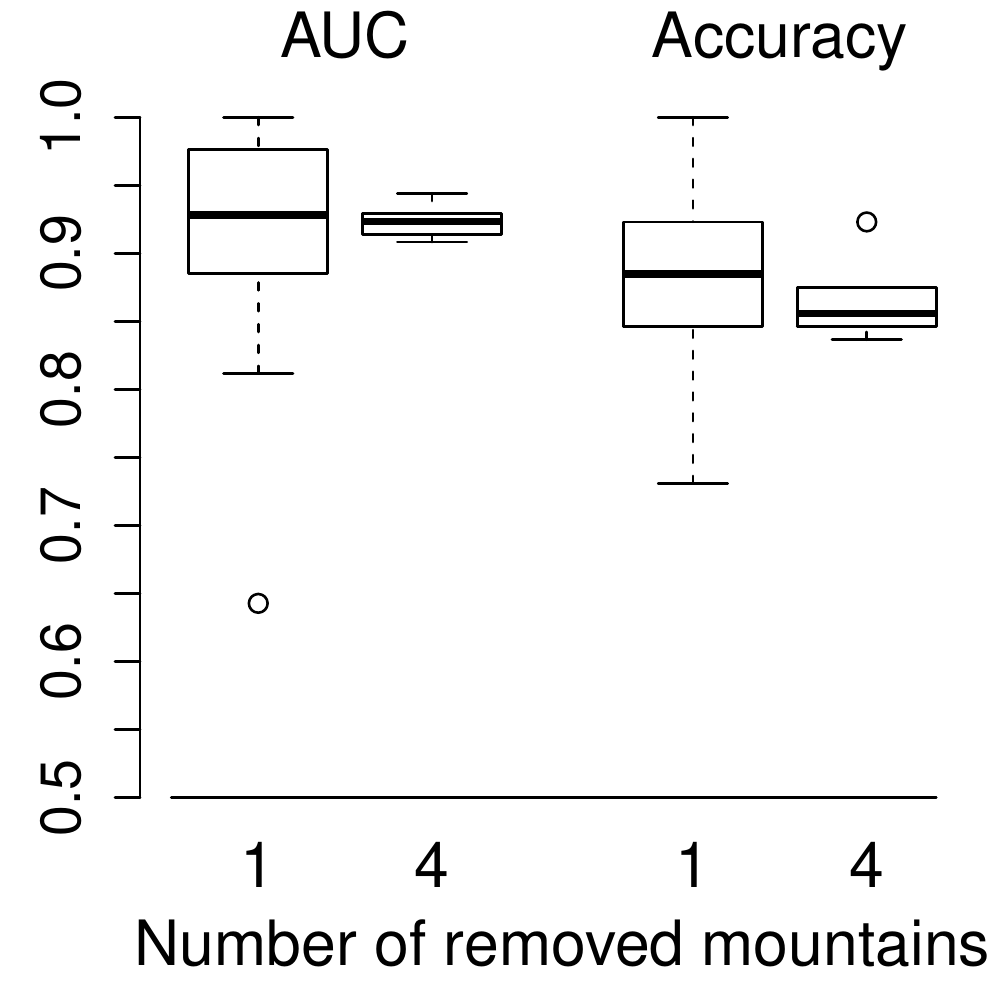}}
	\caption{Performance of the\textit{matching-centrality} model in forecasting the mammal communities on unsampled mountains. We illustrate the performance of the model using an out-of-sample test, removing singly and in groups of four each mountain in the network \cite{Patterson1984}. After fitting the model to the known network, we use a statistical analysis to predict the latent traits of the unsampled mountains from geographical characteristics, and use these traits in the\textit{matching-centrality} model to forecast their mammal community. Graphs show box-plots for the AUC and the accuracy of the forecasts (i.e., the percentage of correctly forecasted 0s and 1s).}
	\label{fig4}
\end{figure}

We illustrate the method using the network describing the presence-absence of mammal species on mountains within the Rocky Mountains \cite{Patterson1984}. The model fits this network perfectly, with an AUC for two \textit{matching} dimension of $1$. Furthermore, we find that the estimated \textit{matching} and \textit{centrality} traits of the mountains are closely related to their geographical characteristics. Specifically, using a generalized least square linear model with a spatial correlation structure, we can relate the latent traits of \textit{matching} and \textit{centrality} to the area, elevation, and geographic position of the mountains (equation (\ref{equ:montains}), Methods, Table S3). Once the model fitted, it is possible to estimate the traits of \textit{matching} and \textit{centrality} for an unsampled mountain based on its geography. Here, because of the correlation structure, the estimation of the latent traits is based on the conditional expectation of a multivariate normal distribution (equation (\ref{equ:centrality_prediction}) and (\ref{equ:matching_prediction}), Methods). The predicted \textit{matching} and \textit{centrality} traits are then used in the \textit{matching-centrality} model (equation (\ref{equ:1})) to estimate the linking probabilities for each mammal species with the unsampled mountain. We demonstrate the capacity of the model by simulation, removing a single mountain or a set of four, and attempting to forecast the mammal fauna. The model performs remarkably well: on average $87\%$ of the data are correctly forecasted (Fig. \ref{fig4}). It must, however, be kept in mind that the applicability of the method depends on the conditions that 1) the \textit{matching-centrality} model fits the network closely, 2) that external information on the nodes can accurately predict the latent traits, and 3) that the nodes to be forecasted must belong to the same statistical population as the observed ones. In all cases, we recommend out-of-sample tests, as explained in the Methods.\\

To our knowledge, only one other method has been devised to forecast the links made by a new node in a network, for a host-parasitoid network based on the phylogenies of the participants \cite{Ives2006}. Our approach has two decisive advantages that make it extremely versatile: firstly, it is not necessary to include information about both sets of nodes of the bipartite network to make the forecast; secondly, by placing latent traits as intermediates between the network adjacency matrix and node characteristics, we provide an entirely flexible way to incorporate external information about nodes, for any conceivable statistical model could be used to relate the latent traits to external variables. Both features are illustrated in the above example, where the forecasting is based only on information about the mountains, and the missing latent traits are estimated from the geographic characteristics using linear predictors and spatial correlation.\\

\section*{Discussion}

By translating an adjacency matrix into a set of quantitative traits for the nodes, the \textit{matching-centrality} model represents a powerful and unifying tool for network analysis.  It allows the reconstruction of missing information and the forecasting of the links of entirely novel nodes, and opens the door to comparative analyses to shed light on the factors underlying network structure across disciplines.\\

\section*{Materials and Methods}

{\bf Fitting of the model}\\
The likelihood of the model is computed by
\begin{equation}
\label{equ:like}
	L = \prod_{ij} P(a_{ij} = 1)^{a_{ij}} (1-P(a_{ij} = 1))^{(1-a_{ij})}
\end{equation}
i.e., we assume the presence-absence of links follows a multi-Bernoulli distribution and that the probabilities are independent given the parameters and the latent traits. As the logit of the linking probabilities is a non-linear function of the parameters, we fit the model using a simulated annealing algorithm \cite{Bremaud1998}.\\

In order to make the parameters and the latent traits of the model uniquely defined we have to impose some constraints:
\begin{enumerate}
	\item All vectors of latent traits $\vec{v^k}, \vec{f^k}, \vec{v^*}, \text{and} \vec{f^*}$ are orthogonal to the unitary vector.
	\item All vector of \textit{matching} trait $\vec{v^k}$ ($k=1,\cdots,d$) are pair-wise orthogonal, and the same for the vectors $\vec{f^k}$.
	\item The length of the vectors $\vec{v^k}, \vec{v^*}$ is set to $\sqrt{(\#S_1)}$, where $\#S_1$ is the number of node in set $S_1$. Similarly for the set $S_2$, the length of the vectors $\vec{f^k}, \vec{f^*}$ is set to $\sqrt{(\#S_2)}$.
\end{enumerate}

{\bf Nodes clustering base on the latent traits}
Using the estimated values of the latent traits of \textit{matching} and \textit{centrality}, we can construct a dendrogram using the following distance $d_{ij}$ between two nodes $i$ and $j$: 
\begin{equation}
\label{equ:distance}
	d_{ij} = \frac{\sqrt{\lambda_1(v_i^1-v_j^1)^2 + \lambda_2(v_i^2-v_j^2)^2}}{\delta v_i^* + \delta v_j^* - 2min_{k}(\delta v_k^*)}
\end{equation}
with the parameters coming from the model (equation (\ref{equ:1})). Note that the numerator is the classical Euclidian distance in a two dimensional \textit{matching} trait space; the denominator is a correction term that weights distances according to the \textit{centrality} of the nodes $i$ and $j$ compared to the minimum \textit{centrality} of the $k$ nodes in the network. The greater the value of the \textit{centrality} trait of a node, the greater the “compression” of the distances; this reflects the fact that the probability of attachment is proportional to the degree of a node (captured by the \textit{centrality} term). \\

{\bf Predicting missing links in partially know networks}\\
We used the terrorist association network \cite{Krebs2002} to evaluate the performance of the \textit{matching-centrality} model to recover missing links. We followed the same procedure as in Clauset et al. \cite{Clauset2008}. Specifically, we simulated networks with missing links by setting a given fraction of 1s to 0s in the adjacency matrix. Then the model is fitted to the incompletely observed networks and latent traits estimated for each node. These \textit{matching} and \textit{centrality} traits are then used to estimate linking probabilities for each pair of nodes (equation \ref{equ:1}). Finally we computed the AUC for the 0's and the missing links pooled. We removed at random 2, 5, 15, 30, 50, 75, 90 and 98\% of the 1's, replicated 100 times for each fraction. For comparison, we present results with the hierarchical model of Clauset et al. \cite{Clauset2008}, block model of Guimera et al. \cite{Guimera2009}, and three classical techniques \cite{Liben-Nowell2007} (Jaccard index, Common neighbors, Degree product).\\

{\bf Linking latent traits and node characteristics}\\
We used phylogenetic regression \cite{Grafen1989} to relate the latent traits of \textit{matching} and \textit{centrality} to species' body-size and phylogeny in the food-web of Tuesday lake. We assume that the latent traits follow a multivariate normal distribution (MVN), where the linear term is given by the logarithm of the body-size and the correlation structure is induced by the phylogeny, i.e.,
\begin{equation}
\label{equ:phylo_regression}
	\vec{v^1},\vec{v^2},\vec{f^1},\vec{f^2},\vec{v^*},\vec{f^*} \sim MVN \left(\alpha + \beta \log(\vec{bs}),\Sigma(\Lambda) \right)
\end{equation}
where: $\vec{v^1}, \vec{v^2}, \vec{f^1}, \vec{f^2}, \vec{v^*}, \vec{f^*}, \vec{bs}$  denote the vectors of the \textit{matching} traits of resources and consumers, the \textit{centrality} traits of resources and consumers, and the body-sizes, respectively; $\Sigma(\Lambda)$ is the variance-covariance matrix; and  $\alpha,\beta$ and $\lambda$ are the parameters of the phylogenetic regression. We use Pagel's-$\lambda$ \cite{Freckleton2002} structure for the variance-covariance matrix, i.e.,
\begin{equation}
\label{equ:phylo_corr}
	\Sigma(\lambda)_{ij} = \left\{ \begin{array}{ccc}
	\sigma^2 \cdot t_{ij} \cdot \lambda & \text{if} & i \neq j \\
	\sigma^2  & \text{if} & i \neq j \end{array}  \right.
\end{equation}
where $\sigma^2$ is the common variance, $t_{ij}$ is the proportion of time that species $i$ and $j$ spent in common before their speciation on the phylogenetic tree, and  is the control parameter for the strength of the phylogenetic correlation ($\lambda=0$ is equivalent to no correlation). The p-values of the parameters $\alpha$ and $\beta$ are computed with the usual z-test, while the p-value associated with the correlation structure is computed using a log-likelihood ratio test between models with and without correlation. The analyses were done in R \cite{Team2009} with the libraries ape \cite{Paradis2004} and nlme \cite{Pinheiro2009}.\\

{\bf Forecasting the links of new nodes}\\
The analyses of the latent traits for the mountains were carried out using generalized least squares regression with a spatial correlation structure \cite{Zuur2009}. We assume that the \textit{matching} and \textit{centrality} traits follow a multivariate normal distribution, where the linear part is given by the longitude, latitude, area and elevation of the mountains, and that the spatial correlation structure is exponential, i.e.,
\begin{equation}
\label{equ:montains}
	\vec{v^1},\vec{v^2},\vec{v^*}  \sim MVN(\alpha + \beta_1 \cdot \text{area} + \beta_2 \cdot \text{elevation} + \beta_3 \cdot \text{latitude} + \beta_4 \cdot \text{longitude},\Sigma(r)), 
\end{equation}
with the elements of the variance-covariance matrix given by
\begin{equation}
\label{equ:spatial_correlation}
	\Sigma(r)_{kl} = \sigma^2 e^{-d_{kj}/r}.
\end{equation}
The vectors $\vec{v^1}, \vec{v^2}, \vec{v^*}$ are the \textit{matching} and \textit{centrality} traits, respectively; $\alpha, \beta_1, \beta_2, \beta_3, \beta_4$ are the intercept and slope parameters; $r$ is a parameter tuning the exponential decay of the spatial correlation; $\sigma$ common variance; and $d_{ij}$ the distance between mountains $i$ and $j$. For the \textit{matching} traits, we found that only the spatial correlation structure was significant (table S3). For the \textit{centrality} traits, all covariates except the longitude and the correlation structure were significant.

Since the \textit{matching} and \textit{centrality} traits of the mountains are significantly related to several covariates and to the correlation structure given by the between-mountain distances, it should be possible to forecast the mammal communities in unsampled mountains for which the covariates are known, given the information provided by the covariates of the sampled mountains and the observed presence/absence network. 

To achieve this task, the first stage is to check that the \textit{matching} and \textit{centrality} traits of the sampled mountains can be recovered from the covariates. If this ability exists, using the model for forecasting is then reasonable. In the first stage, we use conditional expectation to test if there is sufficient predictive power to recover the \textit{matching} and \textit{centrality} traits of the mountains. This out-of-sample test is carried out in the following three steps:
\begin{enumerate}
	\item We refit the model (\ref{equ:montains}) on the data set after removing one mountain, denoted by $k$.
	\item Using the fitted parameters from step 1), we compute the conditional expectation for the \textit{matching} and \textit{centrality} traits of mountain $k$; for the \textit{centrality} trait, this value is given by:	
	\begin{equation}
	\label{equ:centrality_prediction}
		\hat{v_k^*} = \hat{\alpha} + \hat{\beta_1} \cdot \text{area} + \hat{\beta_2} \cdot \text{elevation} + \hat{\beta_3} \cdot \text{latitude}
	\end{equation}	
	where $\hat{\alpha}, \hat{\beta_1}, \hat{\beta_2}, \hat{\beta_3}, \hat{r}$  are the fitted parameters from step 1). The conditional expectation for the \textit{matching} trait is given by:	
	\begin{equation}
	\label{equ:matching_prediction}
		\hat{v_k} = \hat{\alpha} + \Sigma(\hat{r})_{-kk} \Sigma(\hat{r})^{-1}_{kk} \left( \vec{v_{-k}} - \hat{v_{-k}} \right),
	\end{equation}	
	where $\Sigma(\hat{r})_{-kk}$ is the $k$th column without the $k$th row (indicated by subscript $-k$) of the variance-covariance matrix estimated using equation \ref{equ:spatial_correlation}; $\Sigma(\hat{r})_{kk}$ is the $(k,k)$ element of the estimated variance-covariance matrix; $\left( \vec{v_{-k}} - \hat{v_{-k}} \right)$  is the row vector of residuals obtained from step 1. The last term of equation (\ref{equ:matching_prediction}) represents the deviation from the linear prediction that is introduced by knowledge of the spatial correlation structure.	Note that only the spatial correlation and the intercept are significant for the \textit{matching} trait (table S3).
	\item We repeat steps 1 and 2 for all mountains in turn and then finally compare the predicted values of \textit{matching} and \textit{centrality} and the observed values (the values estimated from the \textit{matching-centrality} model fitted to the full network) with a simple linear correlation coefficient.
\end{enumerate}

We found a correlation of $0.95$ for the \textit{centrality} traits, and of $0.71$ and $0.62$ for the two dimensions of \textit{matching} traits. Our model thus has a good predictive power to recover the latent traits of the mountains. This result opens the door to the forecasting of the presence/absence of mammal species on mountains that have not been sampled, but for which we know their characteristics (area, elevation, latitude and distances to the other mountains). In this second stage, we test the forecasting performance of the \textit{matching-centrality} model by removing each mountain from the dataset in turn and attempting to recover its mammal community. This yields the following out-of-sample test:

\begin{enumerate}
	\item	We remove one mountain from the network, and then estimate the \textit{matching} and \textit{centrality} traits of the mammals and of the remaining mountains using the \textit{matching-centrality} model. 
	\item	Based on the model (equation (\ref{equ:montains})) and on the conditional expectation (equations (\ref{equ:centrality_prediction}) and (\ref{equ:matching_prediction})), we predict the \textit{matching} and \textit{centrality} traits for the removed mountain.
	\item	With the predicted \textit{matching} and \textit{centrality} traits, we can predict the linking probabilities (equation \ref{equ:1}) between the removed mountain and the mammals. The presence/absence of the mammals are determined from these linking probabilities and the cut-off point obtained with the fit of step 1. The cut-off point is defined as the linking probability chosen such that the number of false positives is equal to the number of false negatives.
\end{enumerate}

In a real case situation for the forecasting of unsampled nodes, the first condition for application is that the \textit{matching-centrality} model provides a good fit to the sampled network. Secondly, there must exist a relationship between the \textit{matching} and \textit{centrality} traits and the independent information on the nodes; in our example, forecasting the mountains occupied by a new mammal would not be possible. Thirdly, it is necessary that the first stage described above is successful, in our case that the \textit{matching} and \textit{centrality} traits of the sampled mountains can be recovered from the covariates, for otherwise we can have little faith in the results of forecasting. Once these conditions are met, one can proceed to the forecasting of the edges connected to the unsampled nodes. This is performed with steps 2 and 3 of the second stage.\\

We note three final technical points. Firstly, we recommend the performance of a complete out-of-sample test (steps 1 to 3 of the second stage) as additional validation. Secondly, the new nodes have to belong to the same statistical population as the original ones (it would obviously make no sense to forecast the mammal community present in a completely different region). Thirdly, in our example, we related the \textit{matching} and \textit{centrality} traits to the characteristics of the nodes using linear models; our approach is versatile and models of any form can be applied at this stage.\\

{\bf ACKNOWLEDGMENTS} We thank J. Bascompte, S. Saavedra, and A. Davison for critical discussion. The work was funded by the National Centre of Competence in Research "Plant Survival", the Swiss National Science Foundation grant 3100A0-113843 (both to L.-F. Bersier), by SystemsX.ch, the Swiss Initiative in Systems Biology (to L.-F. Bersier and C. Mazza), and by the FP7-REGPOT-2010-1 program (project 264125 EcoGenes) and an ERC Advanced Grant (both to J. Bascompte).\\

\clearpage

%References and Notes
\clearpage
\medskip
\renewcommand{\baselinestretch}{1.5}
{\small
\bibliographystyle{naturemag}
\bibliography{bibliography}
}

\end{document}

\maketitle

\newpage

\renewcommand{\thefigure}{S\arabic{figure}}
\renewcommand{\thetable}{S\arabic{table}}
\renewcommand{\thesection}{S\arabic{section}}
\renewcommand{\theequation}{S\arabic{equation}}
\renewcommand{\figurename}{Figure}
\renewcommand{\tablename}{Table}

\begin{table}[h]
\caption{\textbf{Properties of the seven studied networks and results of model fitting.} For each network, we provide the type of network, the number of nodes ($N$) (for bipartite networks we provide the number of nodes of each type), the number of edges ($E$), the connectance ($c$) (i.e., the fraction of realized links) and, for matching-centrality models with one and two matching dimensions, the area under the ROC-curve (AUC) and the true positive rate ($\Omega$) for the cut-off point such that the number of false positives is equal to the number of false negatives.\label{tbl:S1}}
\resizebox{16cm}{!} {
	\begin{tabular}{p{4cm}ccccccccc}
		\hline
			Network	&	Type	&	$N$	&	$E$	&	$c$	 & \multicolumn{2}{c}{model with $d=1$} &	& \multicolumn{2}{c}{model with $d=2$}\\
			\cline{6-7} \cline{9-10}
			& & & & & AUC & $\Omega$ & & AUC & $\Omega$ \\
		\hline
			Zachary karate club \cite{Zachary1977}	&	undirected	&	34	&	78	&	0.139	&	0.9824	&	0.8205	&		&	0.9996	&	0.9615	\\
			Terrorist association \cite{Krebs2002}	&	undirected	&	62	&	152	&	0.08	&	0.9702	&	0.7237	&		&	0.9977	&	0.9145	\\
			Protein \cite{Maslov2002}	&	undirected	&	159	&	155	&	0.012	&	0.9967	&	0.8000	&		&	1	&	1	\\
			International conflict \cite{Ward2007}	&	directed	&	130	&	203	&	0.012	&	0.9845	&	0.6798	&		&	0.9982	&	0.8916	\\
			Food web \cite{Jonsson2005}	&	directed	&	66	&	377	&	0.087	&	0.9948	&	0.9204	&		&	1	&	1	\\
			Seed dispersal web \cite{Rezende2007}	&	bipartite	&	79/25	&	299	&	0.151	&	0.9756	&	0.8571	&		&	0.9977	&	0.9610	\\
			Mammals-mountains \cite{Patterson1984}	&	bipartite	&	28/26	&	275	&	0.378	&	0.9963	&	0.9636	&		&	1	&	1\\	
		\hline
	\end{tabular}
}
\end{table}

\clearpage
\section*{Association between terrorists \cite{Krebs2002}}

\begin{figure}[h]
	\centerline{\includegraphics*[width= 0.9 \linewidth]{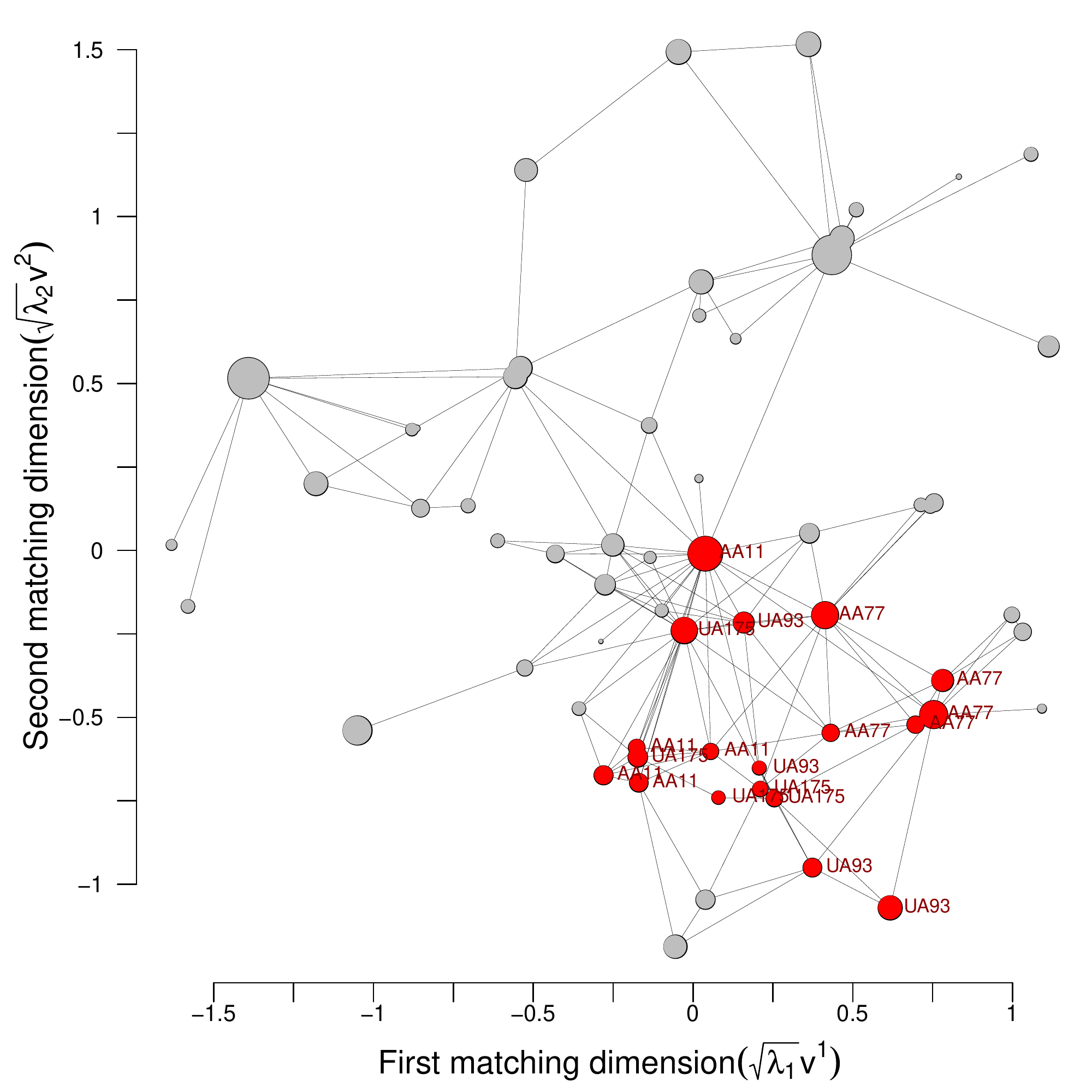}}
	\caption{\textit{Matching} traits space representation of a network of associations between terrorists involved in the September 11 2001 terrorist attacks in the US. The size of each node is proportional to its \textit{centrality} trait value. Labels next to the nodes give the flight hijacked by the terrorist.}
	\label{figS1}
\end{figure}

This network describes the links between 62 individuals that were directly and indirectly involved in the September 11 2001 terrorist attacks in the US. It is an expanded version of the network shown in Fig. 4 in Krebs \cite{Krebs2002}, available here \cite{Clauset}. The two-dimensional latent space representation clearly exhibits two clusters of nodes, which are connected by a central terrorists that was on flights AA11 (Fig. \ref{figS1}).

\clearpage
\section*{Protein interactions \cite{Maslov2002}}

\begin{figure}[h]
	\centerline{\includegraphics[width= 0.9 \linewidth]{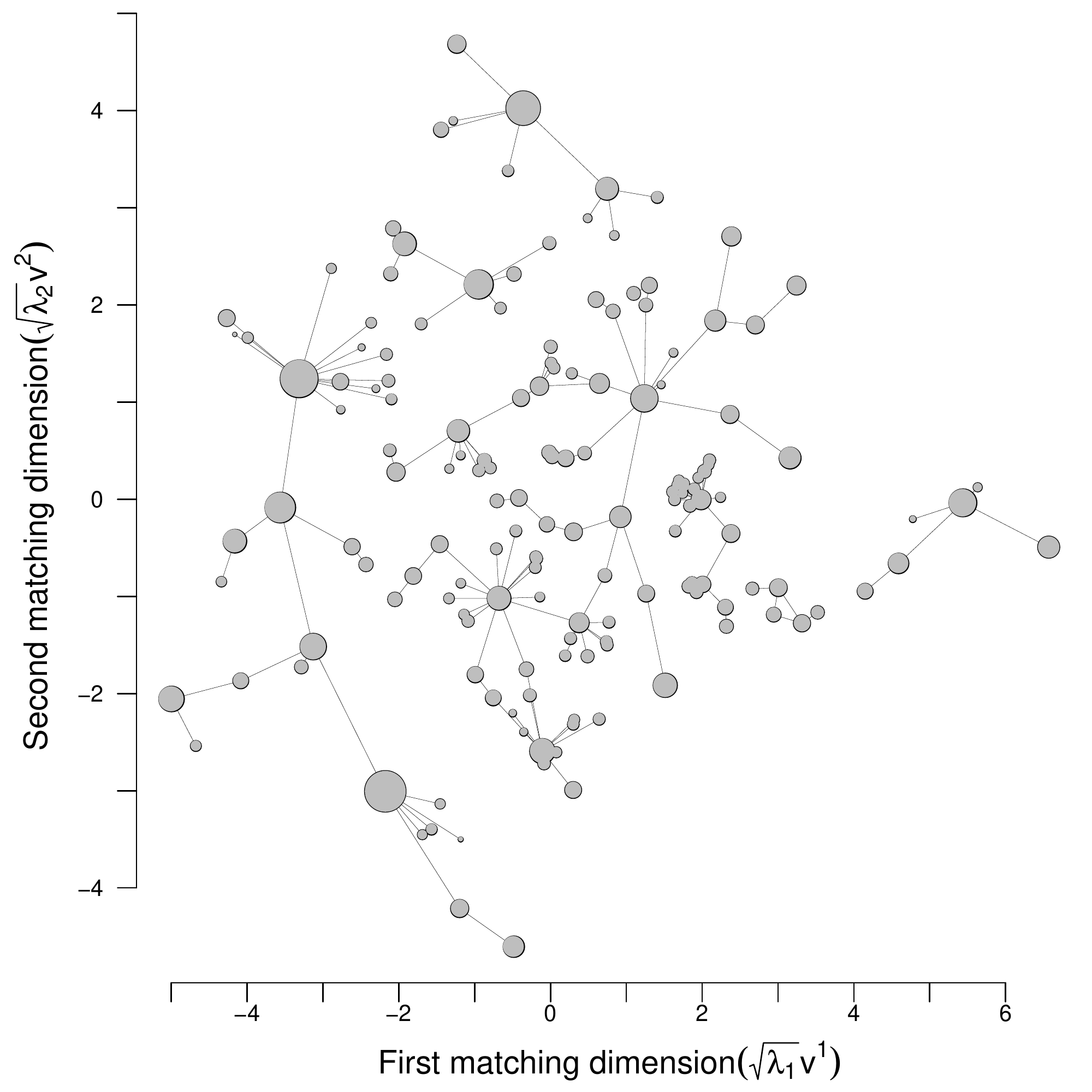}}
	\caption{\textit{Matching} traits space representation of a protein network. The size of each node is proportional to its \textit{centrality} trait value.}
	\label{figS2}
\end{figure}

This data is a subset of the protein interaction network in Saccharomyces cerevisiae, involving proteins that are localized within the nucleus and that interact with at least one other nuclear protein \cite{Melian2002} (Fig. \ref{figS2}).

\clearpage
\section*{International conflicts between countries \cite{Ward2007}}

\begin{figure}[h]
	\centerline{\includegraphics[width= 0.8 \linewidth]{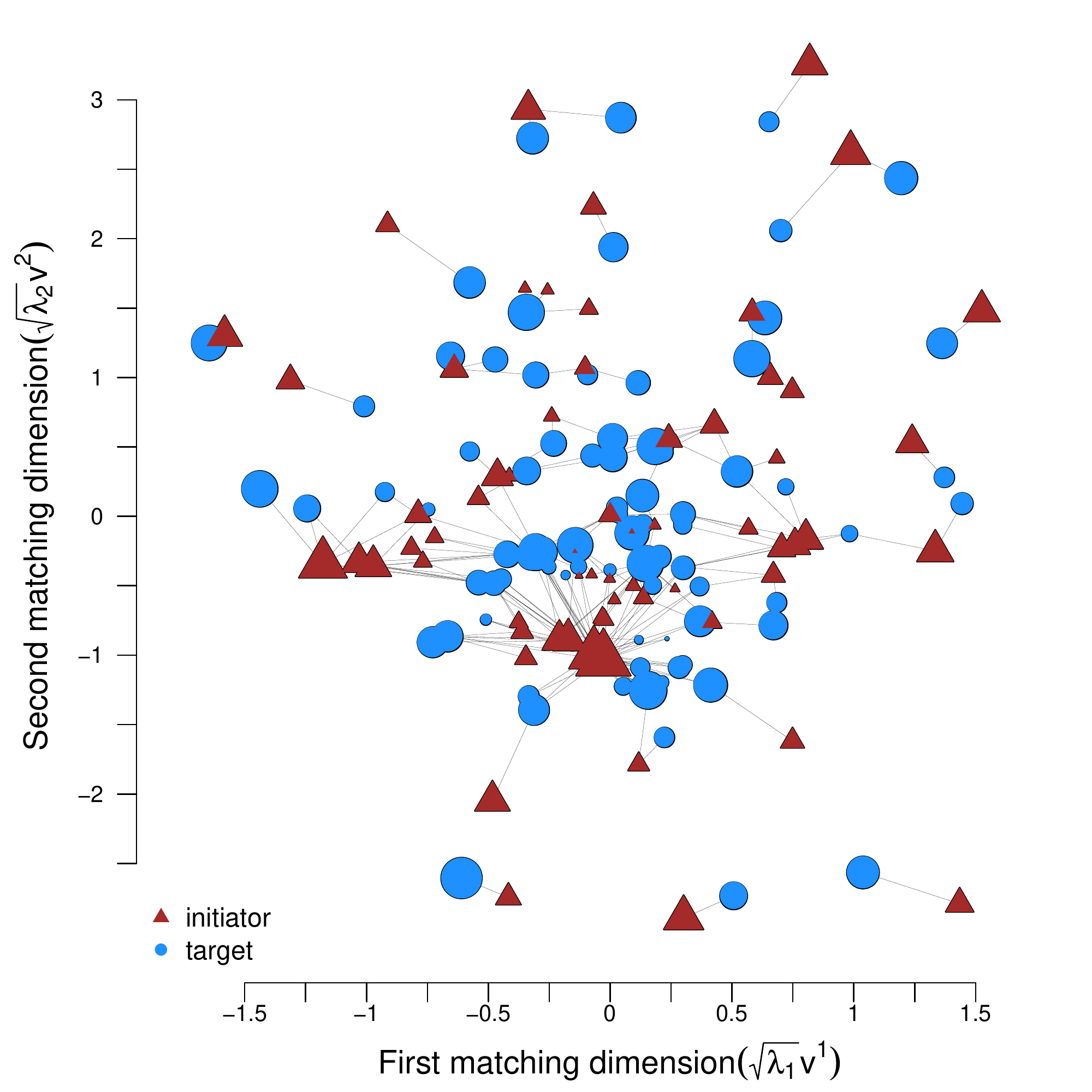}}
	\caption{\textit{Matching} traits space representation of the international conflict network between countries for the period 1990-2000. The size of each node is proportional to its \textit{centrality} trait value. Each country is represented twice, as initiator and as target.} 
	\label{figS3}
\end{figure}

This dataset consists of 203 international military conflicts between 130 countries during the period 1990-2000. It forms a directed network, where the adjacency matrix is given by $a_{ij} = 1$ if country $i$ initiates conflict with country $j$. The latent traits space representation (Fig. \ref{figS3}) clearly exhibits a large clusters of nodes and some unconnected clusters made of few countries.

\clearpage
\section*{Food web of Tuesday Lake \cite{Jonsson2005}}

\begin{figure}[h]
	\centerline{\includegraphics[width= 0.9 \linewidth]{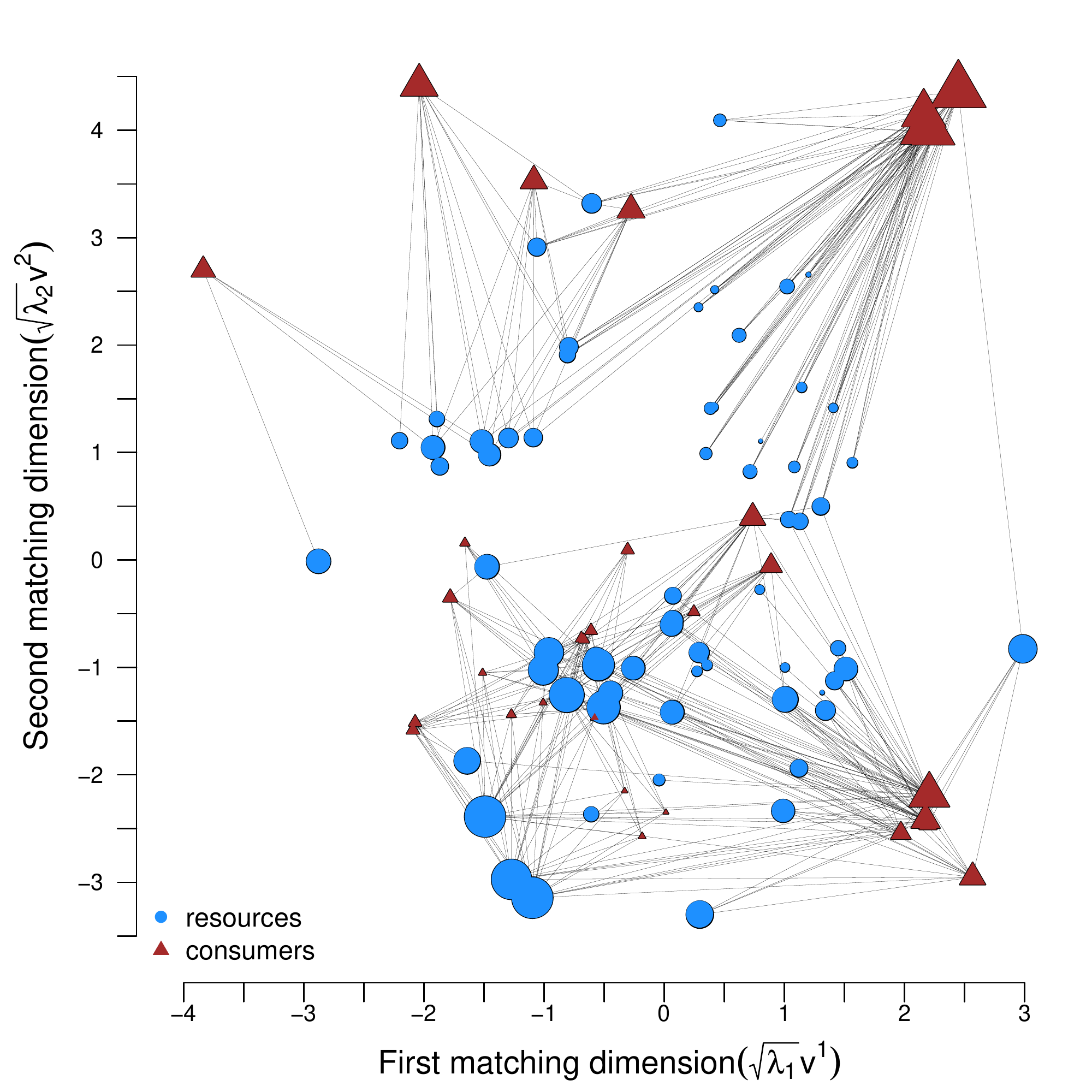}}
	\caption{\textit{Matching} traits space representation of the Tuesday Lake food web. Each species is represented twice, once in its role as consumer, and once in its role as resource. The size of each node is proportional to its \textit{centrality} trait value.}
	\label{figS4}
\end{figure}

This network describes the trophic links (who eats whom) between 66 species in Tuesday Lake (Michigan, USA). It forms a directed network, where the adjacency matrix is given by $a_{ij} = 1$ if species $i$ is eaten by species $j$. We remark that one matching dimension is able to fit correctly 92\% of the links, while two dimensions explain perfectly the network (Table \ref{tbl:S1}, Fig. \ref{figS4}).

\begin{table}[h]
\caption{\textbf{Results of the phylogenetic regressions for the Tuesday Lake food web.} We provide the results of the phylogenetic regressions for the centrality and matching traits. For the body size covariate we provide the estimated slope ($\beta$) and its p-value computed by the z-test. For the phylogeny we provide Pagel's $\lambda$ (i.e., the strength of the phylogenetic signal) and its p-value computed using a log-likelihood ratio test between models with and without the phylogenetic correlation. Response variables and the body-size covariate were standardized. \label{tbl:S3}}
\centering
\resizebox{14cm}{!}{
	\begin{tabular}{p{5cm}ccccc}
		\hline
			& \multicolumn{2}{c}{Body size} & & \multicolumn{2}{c}{Phylogeny} \\ \cline{2-3} \cline{5-6} 
			&	Parameter ($\beta$) 	&	p-value	&		&	Parameter ($\lambda$)	&	p-value	\\
		\hline
			Resources	&	&	&	&  & \\ \cline{1-1}
			First matching dimension	&	\textbf{-0.066}	&	\textbf{0.026}	&		&	\textbf{0.438} &	\textbf{0.001}	\\
			Second matching dimension	&	\textbf{0.095}	&	\textbf{0.010}	&		&	\textbf{0.694}	&	\textbf{$<$0.001} \\
			Centrality traits	&	\textbf{-0.009}	&	\textbf{0.008}	&		&	\textbf{0.560}	&	\textbf{0.001} \\
			& & & & & \\
			Consumers	&	&	&	&  & \\ \cline{1-1}
			First matching dimension	&	\textbf{-0.137}	&	\textbf{0.049}	&		&	\textbf{0.802} &	\textbf{$<$0.001} \\
			Second matching dimension	&	\textbf{0.222}	&	\textbf{0.039}	&		&	\textbf{0.844}	&	\textbf{0.001} \\
			Centrality traits	&	\textit{0.014}	&	\textit{0.054}	&		&	\textbf{0.923}	&	\textbf{$<$0.001} \\
		\hline
	\end{tabular}
}
\end{table}

\clearpage
\section*{Mutualistic seed-dispersal network of Nava de las Correhuelas \cite{Rezende2007}}

This network describes the mutualistic links between 25 plant and 33 bird species of Nava de las Correhuelas (Sierra de Cazorla, southern Spain). It is a bipartite network, with the two sets of nodes formed by plants and birds. The adjacency matrix is given by $a_{ij} = 1$ if bird $i$ feeds on the fruits, and then disperses the seeds, of plant $j$. 

\begin{figure}[h]
	\centerline{\includegraphics[width= 0.8 \linewidth]{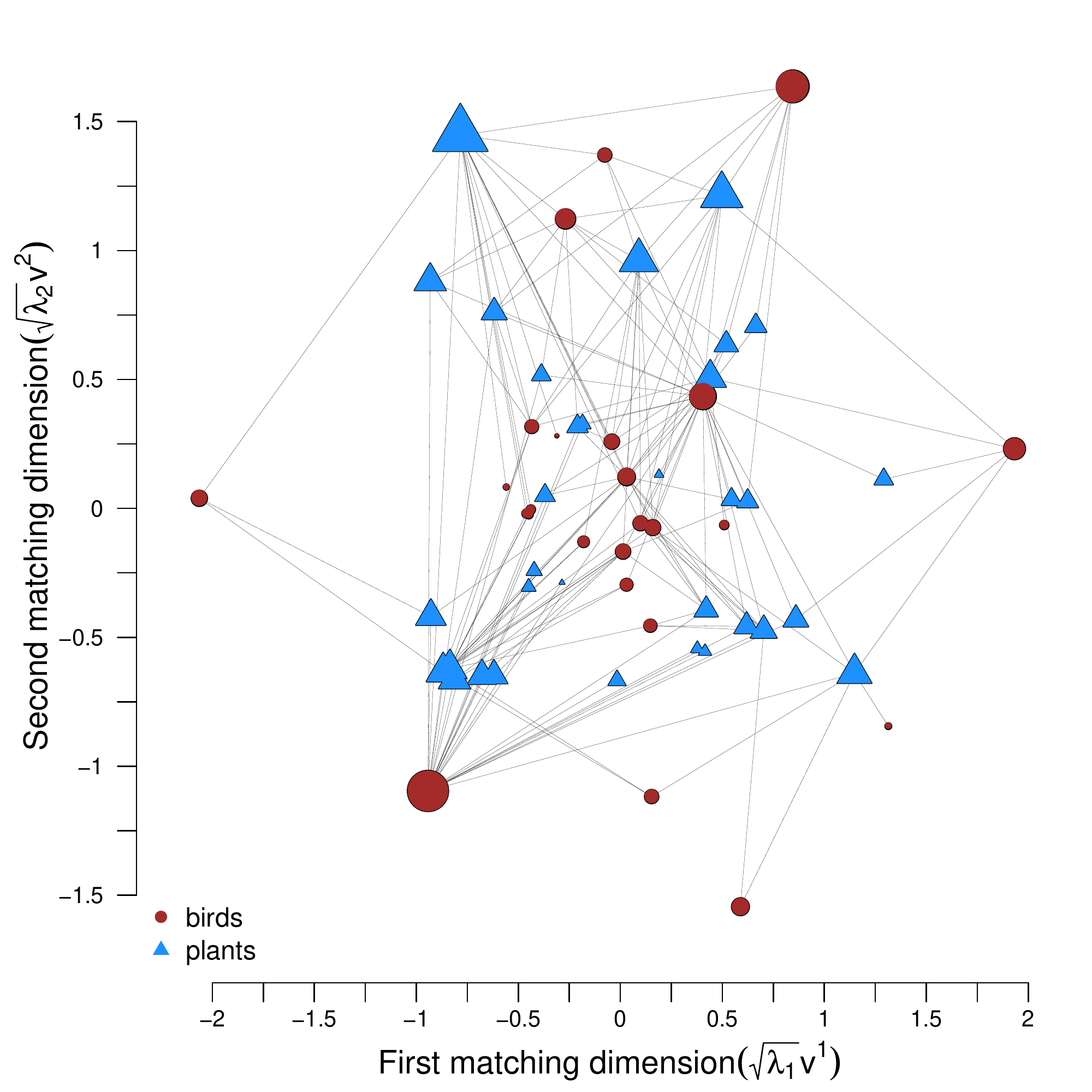}}
	\caption{\textit{Matching} traits space representation of the mutualistic seed-dispersal web of Nava de las Correhuelas. The size of each node is proportional to its \textit{centrality} trait value.}
	\label{figS5}
\end{figure}

\clearpage
\section*{Presence-absence data of mammal communities on mountains \cite{Patterson1984}}

\begin{figure}[h]
	\centerline{\includegraphics[width= 0.85 \linewidth]{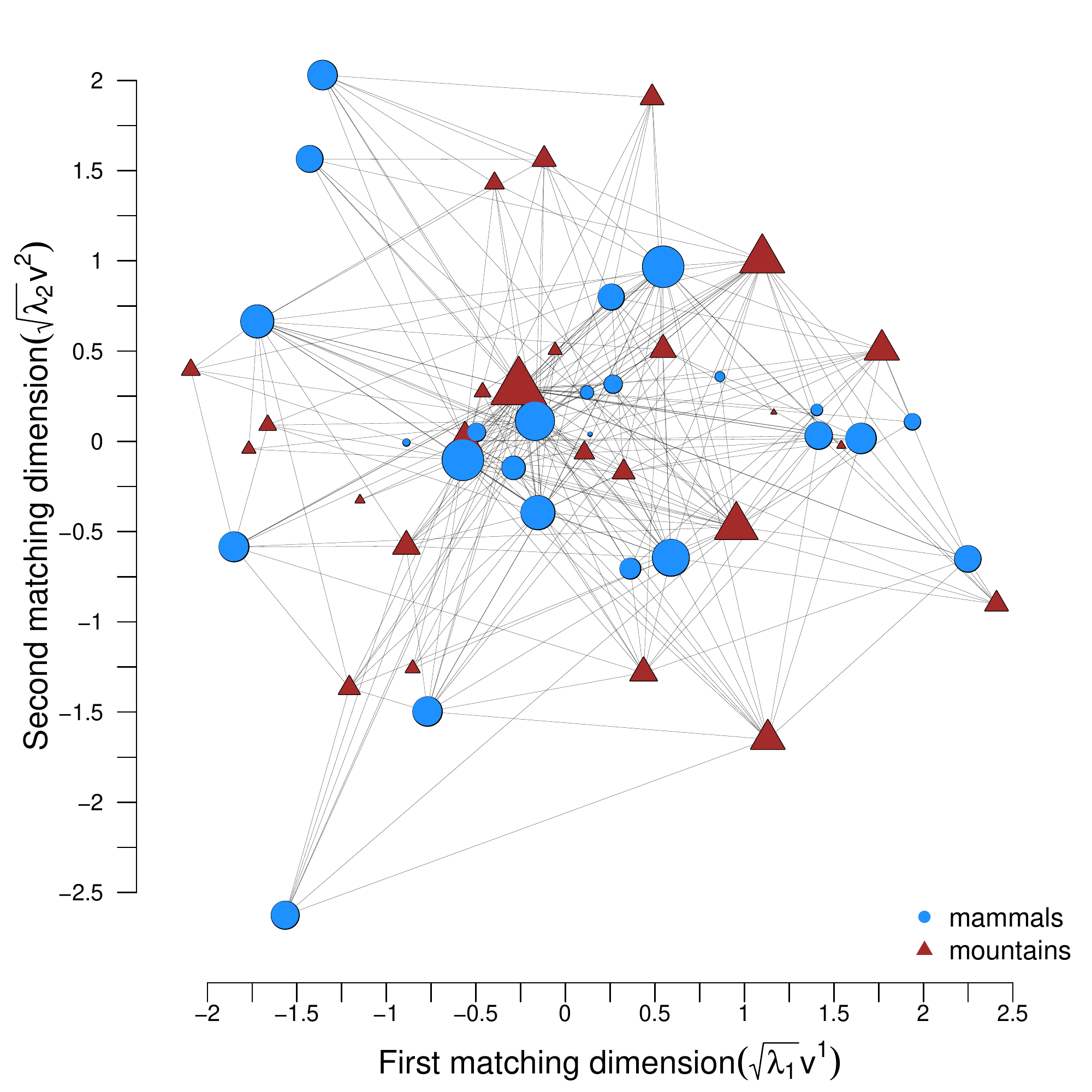}}
	\caption{\textit{Matching} traits space representation of the mammals-mountains network. The size of each node is proportional to its \textit{centrality} trait value.}
	\label{figS6}
\end{figure}

This dataset describes the distribution of 26 mammal species on 28 mountains within the southern Rocky Mountains (USA). It can be considered as a bipartite network, with nodes in set $S_1$ representing the mountains and those in set $S_2$ the mammal species, and the adjacency matrix is then given by $a_{ij} = 1$ if species $j$ is present on mountain $i$ (Fig. \ref{figS6}). 

\begin{table}[h]
\centering
\caption{\textbf{Results of phylogenetic regression and regression with spatial correlation on the latent traits from the mammals-mountains network.} We provide the results of the phylogenetic regression for the mammals and the generalized least squares regression with spatial correlation structure for the mountains. \label{tbl:S4}}
\resizebox{15cm}{!} {
	\begin{tabular}{p{3.5cm}cccccccc}
		\hline
			Covariate & \multicolumn{2}{c}{Centrality traits} & & \multicolumn{2}{c}{First matching dimension} & & \multicolumn{2}{c}{Second matching dimension}\\ \cline{2-3} \cline{5-6} \cline{8-9}
			&	Parameter	&	p-value	&		&	Parameter	&	p-value &		&	Parameter	&	p-value		\\
		\hline
			For mammals & & & &	& &	& &	\\ \cline{1-1}
			Body-size	&	-	&	n.s.	&		&	0.399	&	0.044 & & - & n.s. \\
			Phylogeny	&	-	&	n.s.	&		&	-	&	n.s.& & -  &	 n.s.\\
			&		&		&		&		&	& & &	\\
			For mountains	&		&		&		&		& & & &	\\ \cline{1-1}
			Longitude	&	-	&	n.s. &		&	-	&	n.s. & & - & n.s. \\
			Latitude	&	0.4955	&	$<$0.001	&		&	-	&	n.s. & & - & n.s. \\
			Area	&	0.2147	&	0.016	&		&	-	&	n.s. & & - & n.s. \\
			Elevation	&	0.4360	&	$<$0.001	&		&	-	&	n.s. & & - & n.s. \\
			Distance	&	-	&	n.s. &		&	0.013	&	$<$0.001 & & 0.0079 & $<$0.001\\
		\hline
	\end{tabular}
}
\end{table}

\clearpage
\medskip

\renewcommand{\baselinestretch}{1.5}
{\small
\bibliographystyle{naturemag}
\bibliography{bibliography}

}